\documentclass[preprint,pra,showpacs,nofootinbib]{revtex4}
\usepackage{float,epsfig}
\usepackage{bm}
\usepackage{graphicx}
\usepackage{subfigure}
\usepackage{dcolumn}
\usepackage{amsmath,amssymb,amsthm}
\usepackage[toc,page,title,titletoc,header]{appendix}
\usepackage{bm}
\def\/{\over}
\begin{document}
\title{\bf Position dependent energy  level shifts of an accelerated atom in the presence of a boundary}
\author{ Zhiying Zhu$^{1,2}$ and  Hongwei Yu$^{1,3}$\footnote{Corresponding author} }
\affiliation{ $^1$Department of Physics  and Key Laboratory of Low
Dimensional
Quantum Structures and Quantum Control of Ministry of Education,\\
Hunan Normal University, Changsha, Hunan 410081, China\\
$^2 $Department of Physics and Electronic Science, Changsha
University of Science and Technology, Changsha, Hunan 410076,
China\\
$^3$ Center for Non-linear Science and Department of Physics, Ningbo
University, Ningbo, Zhejiang, 315211
 China}


\begin{abstract}
We consider a uniformly accelerated atom interacting with a vacuum
electromagnetic field in the presence of an infinite conducting
plane boundary and calculate separately the contributions of vacuum
fluctuations and radiation reaction to the atomic energy  level
shift. We analyze in detail the behavior of the total energy shift
in three different regimes of the distance in both the low
acceleration and high acceleration limits. Our results show that, in
general, an accelerated atom does not behave as if immersed in a
thermal bath at the Unruh temperature in terms of the atomic energy
level shifts, and the effect of the acceleration on the atomic
energy level shifts may in principle become appreciable in certain
circumstances, although it may not be realistic for actual
experimental measurements. We also examine the effects of the
acceleration on the level shifts when the acceleration is of the
order of the transition frequency of the atom and we find some
features  differ from what was obtained in the existing literature.

\end{abstract}

\maketitle

\section{Introduction}

An important prediction of quantum field theory is the existence of
quantum fluctuations of electromagnetic fields even in vacuum. These
quantum fluctuations lead to a number of observable effects such as
the Lamb shift, and the Casimir and Casimir-Polder forces (see,
Ref.~\cite{BKMM} for an extensive review). All these effects have
been observed experimentally. The Casimir-Polder force was
originally studied between a neutral electric polarizable atom at
rest and a conducting plane in vacuum by Casimir and Polder
\cite{CP}, and later a variety of situations have been considered,
for example, the case of an atom near the surface of a dielectric
slab \cite{CRE} or a nanostructure \cite{Yannopapas}, and the case
of the atom-surface system  in or out of thermal equilibrium
\cite{Zhu-Yu09, Antezza,Buhmann}. In addition, the case of atoms on
accelerated trajectories has also been investigated
\cite{Audretsch95, Hu,Passante97,Rizzuto07,Rizzuto09}. Our interest
in the Casimir-Polder force associated with accelerated atoms is
two-fold. First, such studies may shed some light on our
understanding of the Unruh effect~\cite{FDU}, on which controversy
still exists~\cite{Controversy}, and, second,  the CP force of an
accelerated atom, which arises as a result of the change of atomic
energy level shifts, may provide a new possibility to detect the
Unruh effect.

In this regard, let us note that the Casimir-Polder interaction
energy between an accelerated atom in interaction with the vacuum
electromagnetic fields and an infinite conducting plane boundary,
which is equal to the energy shift of the atom caused by the
presence of the boundary, has recently been studied, and it is found
that the effect of acceleration is not purely
thermal~\cite{Rizzuto09}. In that work, a generalization of the
formalism suggested by Dalibard, Dupont-Roc and
Cohen-Tannoudji~\cite{Dalibard82,Dalibard84}(DDC), which allows a
separate calculation of contributions of vacuum fluctuation and
radiation reaction to the energy level shift is employed, and the
result is however expressed as an integral of a very complicated
function  which makes it  hard both for analytical examination of
the behavior of the level shifts in different distance regimes and
for numerical analysis. In the present paper, we plan to revisit the
problem. We have been able to obtain a result of the energy level
shift of the accelerated atom, which is almost in a closed form
except for a part which is a integral of a much simpler function.
This enables us to make a thorough comparison, in different distance
regimes, with the result of the case of a static atom in a thermal
bath found by us using the same formalism~\cite{Zhu-Yu09} to see
whether the accelerated atom behaves as if immersed in a thermal
bath in terms of atomic energy level shifts~\footnote{Let us note
that it has been demonstrated that the accelerated atom near a
boundary behaves differently from the inertial one immersed in a
thermal bath in terms of the spontaneous excitation
rate~\cite{Audretsch94,Yu-Lu05,Zhu-Yu06,Yu-Zhu06,Zhu-Yu07}.}. More
importantly, the numerical analysis of our result at an acceleration
of the transition frequency of the atom yields conclusions that
differ from those obtained in Ref.~\cite{Rizzuto09}. For example, we
find, contrary to the conclusion in Ref.~\cite{Rizzuto09}, that the
energy level shift of an atom with an acceleration of the typical
transition frequency in the vicinity of an infinite conducting plane
is smaller than that of a static one when the distance is at the
order of $10^{-6}$m, among others.

The paper is organized as follows, in Sec.~II, using the DDC
formalism, we separately calculate both the contributions of vacuum
fluctuations and radiation reaction to the position dependent energy
shifts of an atom on a uniformly accelerated trajectory, which gives
rise to the Casimir-Polder force on the accelerated atom. In
Sec.~III, we analyze in detail the behavior of the energy shift of a
ground-state atom in three different regimes of the distance in both
the low acceleration and high acceleration limits. In addition, with
the method of numerical analysis, we also discuss the behavior of
the energy shift for an atom with a typical acceleration necessary
to observe the Unruh effect. Comparing our results with those of a
static atom immersed in a thermal bath, we can see explicitly the
difference between the energy shift of an accelerated atom and that
of a thermal one. Finally, we will conclude in Sec.~IV with a
summary of results obtained.

\section{vacuum fluctuations and radiation reaction contributions to the energy shifts of an accelerated atom in the presence of a boundary}
We consider a uniformly accelerated two-level atom in interaction with the
vacuum electromagnetic fields in a flat spacetime with an infinite
conducting plane boundary.  Let us note that for a fully realistic
treatment, one may need to consider a multilevel atom.  Using the
DDC formalism and following the procedures that have been shown in
Refs.~\cite{Dalibard84,Passante97,Zhu-Yu09,Rizzuto09}, we will
calculate separately the contributions of vacuum fluctuations and
radiation reaction to the energy shifts of a two-level atom. To be
self-contained, we will first review the general formalism.
Employing the Hamiltonian of the atom-field system that has been
given in Ref~\cite{Zhu-Yu09}, one can write down the Heisenberg
equations of motion for the dynamical variables of the atom and
field. The solutions of the equations of motion can be split into
the two parts: a free part, which is present even in the absence of
the coupling, and which we will denote with the superscript $f$, and
a source part, which is caused by the interaction of the atom and
field, and which we will denote with the superscript $s$. We assume
that the initial state of the field is the Minkowski vacuum
$|0\rangle$, while the atom is in the state $|b\rangle$. To identify
the contributions of vacuum fluctuations and radiation reaction to
the radiative energy shifts of our accelerated atoms, we choose a
symmetric ordering between the atom and the electric field variables
and consider the effects of $E^f$ (corresponding to the effect of
vacuum fluctuations) and $E^s$ (corresponding to the effect of
radiation reaction) separately in the Heisenberg equations of an
arbitrary atomic observable. Then taking the average in the vacuum
state of the electromagnetic field, one can obtain the effective
Hamiltonians
\begin{eqnarray}
H^{eff}_{vf}(\tau)=-{i\/2}\int^\tau_{\tau_0}d\tau'C^F_{ij}(x(\tau),x(\tau'))[\mu_i^f(\tau),\mu_j^f(\tau')]\;,\label{Hvf}
\end{eqnarray}
\begin{eqnarray}
H^{eff}_{rr}(\tau)=-{i\/2}\int^\tau_{\tau_0}d\tau'\chi^F_{ij}(x(\tau),x(\tau'))\{\mu_i^f(\tau),\mu_j^f(\tau')\}\;,\label{Hrr}
\end{eqnarray}
where $\mu_i$ ($\mu_j$) is a component of the atomic electric dipole
moment, $\tau$ is the proper time, and $x(\tau)$ is the stationary
trajectory of the atom. Here we take a perturbation treatment up to
order $\mu^2$, and use $[\ ,\ ]$ and $\{\ ,\ \}$ to denote the
commutator and anticommutator. The subscript ``$vf$" and ``$rr$"
stand respectively for the contributions of vacuum fluctuations and
radiation reaction. The statistical functions $C^F_{ij}$ and
$\chi^F_{ij}$ are defined as
\begin{eqnarray}
C^F_{ij}(x(\tau),x(\tau'))={1\/2}\langle0|\{E_i^f(x(\tau)),E_j^f(x(\tau'))\}|0\rangle\;,\label{cf}
\end{eqnarray}
\begin{eqnarray}
\chi^F_{ij}(x(\tau),x(\tau'))={1\/2}\langle0|[E_i^f(x(\tau)),E_j^f(x(\tau'))]|0\rangle\;,\label{xf}
\end{eqnarray}
which are also called the symmetric correlation function and the
linear susceptibility of the field. Taking the expectation value of
Eqs.~(\ref{Hvf}) and (\ref{Hrr}) in the atom's initial state
$|b\rangle$, we can obtain the vacuum fluctuations and radiation
reaction contributions to the radiative energy shifts of the atom's
level $|b\rangle$,
\begin{eqnarray}
(\delta
E_b)_{vf}=-i\int^\tau_{\tau_0}d\tau'C^F_{ij}(x(\tau),x(\tau'))(\chi^A_{ij})_b(\tau,\tau')\;,\label{Evf}
\end{eqnarray}
\begin{eqnarray}
(\delta
E_b)_{rr}=-i\int^\tau_{\tau_0}d\tau'\chi^F_{ij}(x(\tau),x(\tau'))(C^A_{ij})_b(\tau,\tau')\;,\label{Err}
\end{eqnarray}
where  $(C^A_{ij})_b$ and $(\chi^A_{ij})_b$, the symmetric
correlation function and the linear susceptibility of the atom, are
defined as
\begin{eqnarray}
(C^A_{ij})_b(\tau,\tau')={1\/2}\langle
b|\{\mu_i^f(\tau),\mu_j^f(\tau')\}|b\rangle\;,
\end{eqnarray}
\begin{eqnarray}
(\chi^A_{ij})_b(\tau,\tau')={1\/2}\langle
b|[\mu_i^f(\tau),\mu_j^f(\tau')]|b\rangle\;,
\end{eqnarray}
which are just characterized by the atom itself. For the explicit
forms, the statistical functions of the atom can be written as
\begin{eqnarray}
(C_{ij}^A)_b(\tau,\tau')={1\/2}\sum_d[\langle
b|\mu_i(0)|d\rangle\langle d|\mu_j(0)|b\rangle
e^{i\omega_{bd}(\tau-\tau')}+\langle b|\mu_j(0)|d\rangle\langle
d|\mu_i(0)|b\rangle e^{-i\omega_{bd}(\tau-\tau')}]\label{ca}\;,\nonumber\\
\end{eqnarray}\begin{eqnarray}
(\chi_{ij}^A)_b(\tau,\tau')={1\/2}\sum_d[\langle
b|\mu_i(0)|d\rangle\langle d|\mu_j(0)|b\rangle
e^{i\omega_{bd}(\tau-\tau')}-\langle b|\mu_j(0)|d\rangle\langle
d|\mu_i(0)|b\rangle e^{-i\omega_{bd}(\tau-\tau')}]\label{xa}\;,\nonumber\\
\end{eqnarray}
where $\omega_{bd}=\omega_b-\omega_d$ and the sum extends over a
complete set of atomic states.

In order to calculate the statistical functions of the electric
field, Eqs.~(\ref{Evf}) and (\ref{Err}), we will firstly consider
the two point function of the four potential, $A^\mu(x)$, which can
be obtained by the method of images. At a distance $z$ from the
boundary, we find in the laboratory frame,
\begin{eqnarray}
D^{\mu\nu}(x,x')=\langle0|A^\mu(x)A^\nu(x')|0\rangle=D_{free}^{\mu\nu}(x-x')+D_{bnd}^{\mu\nu}(x,x')\label{D}\;,
\end{eqnarray}
where
\begin{eqnarray}
D_{free}^{\mu\nu}(x-x')={\eta^{\mu\nu}\/{4\pi^2
[(t-t^\prime-i\varepsilon)^2-(x-x^\prime)^2-(y-y^\prime)^2-(z-z^\prime)^2]}}\;,
\end{eqnarray}
is the two point function in the free space and
\begin{eqnarray}
D_{bnd}^{\mu\nu}(x,x')=-{{\eta^{\mu\nu}+2n^\mu n^\nu}\/{4\pi^2
[(t-t^\prime-i\varepsilon)^2-(x-x^\prime)^2-(y-y^\prime)^2-(z+z^\prime)^2]}}\;,\label{Dbound}
\end{eqnarray}
represents the correction induced by the presence of the conducting
boundary. Here $\varepsilon\rightarrow+0$,
$\eta^{\mu\nu}$=diag$(1,-1,-1,-1)$, the unit normal vector
$n^\mu=(0,0,0,1)$, and the subscript ``$bnd$" stands for the part
induced by the presence of the boundary. From Eq.~(\ref{D}), we can
get the electric field two point function in the laboratory frame,
\begin{eqnarray}
\langle0|E_i(x(\tau))E_j(x(\tau'))|0\rangle=\langle0|E_i(x(\tau))E_j(x(\tau'))|0\rangle_{free}
+\langle0|E_i(x(\tau))E_j(x(\tau'))|0\rangle_{bnd}\;,\label{EE}
\end{eqnarray}
where
\begin{eqnarray}
\langle0|E_i(x(\tau))E_j(x(\tau'))|0\rangle_{free}&=&{1\/4\pi^2}(\delta_{ij}\partial
_0\partial_0^\prime-\partial_i\partial_j^\prime)\nonumber\\&&\times
{1\/(x-x')^2+(y-y')^2+(z-z')^2-(t-t'-i\varepsilon)^2}\;,
\end{eqnarray}
and
\begin{eqnarray}
\langle0|
E_i(x(\tau))E_j(x(\tau'))|0\rangle_{bnd}&=&-{1\/4\pi^2}[\,(\delta_{ij}-2n_in_j)\,\partial
_0\partial_0^\prime-\partial_i\partial_j^\prime\,]\nonumber\\&&\times
{1\/(x-x')^2+(y-y')^2+(z+z')^2-(t-t'-i\varepsilon)^2}\;.\label{EEbound}
\end{eqnarray}
Here $\partial^\prime$ denotes the differentiation with respect to
$x^\prime$. Then the statistical functions of the electric field can
be found from Eq.~(\ref{EE}) as a sum of the free space part and the
boundary-dependent part.

In the present paper, we are interested in the energy shifts of
atomic levels caused by the presence of the plane boundary. These
energy level shifts are position dependent and give rise to the
Casimir-Polder force acting on the atom.  So, subtracting the free
space part in Eqs.~(\ref{Evf}) and (\ref{Err}), we can obtain the
contributions of vacuum fluctuations and radiation reaction to the
boundary-dependent energy shifts of atomic levels,
\begin{eqnarray}
(\delta
E_b)_{vf}^{bnd}=-i\int^\tau_{\tau_0}d\tau'(C^F_{ij})_{bnd}(x(\tau),x(\tau'))(\chi^A_{ij})_b(\tau,\tau')\;,\label{Ecpvf}
\end{eqnarray}
\begin{eqnarray}
(\delta
E_b)_{rr}^{bnd}=-i\int^\tau_{\tau_0}d\tau'(\chi^F_{ij})_{bnd}(x(\tau),x(\tau'))(C^A_{ij})_b(\tau,\tau')\;.\label{Ecprr}
\end{eqnarray}
Therefore the total energy shifts caused by the presence of a
conducting plane boundary is
\begin{eqnarray}
(\delta E_b)_{tot}^{bnd}=(\delta E_b)_{vf}^{bnd}+(\delta
E_b)_{rr}^{bnd}\;.
\end{eqnarray}

We assume that the conducting plane boundary is located at $z=0$ in
space, and the two-level atom is being uniformly accelerated in the
$x$-direction with a proper acceleration $a$ at a distance $z$ from
the boundary. The atom's trajectory can be described with the proper
time $\tau$ by
\begin{equation}
t(\tau)={1\/a}\sinh a\tau\;,\ \ \ x(\tau)={1\/a}\cosh a\tau\;
,\ \ \ y(\tau)=y_0\;,\ \ \ z(\tau)=z\;.\label{acctraj}
\end{equation}
With this trajectory, we can calculate the boundary-dependent two
point function of the electric field in the proper reference frame
of the atom with a Lorentz transformation in Eq.~(\ref{EEbound}),
\begin{eqnarray}
\langle0|E_i(x(\tau))E_j(x(\tau'))|0\rangle_{bnd}
&=&-{a^4\/16\pi^2}{1\/
[\,\sinh^2{a\/2}(u-i\varepsilon)-a^2z^2\,]^3}\nonumber\\
&& \times\bigg\{\big[\,\delta_{ij}-2n_in_j+2az(n_ik_j+k_in_j)
\,\big]\sinh^2{au\/2}\nonumber\\
&& \quad+
a^2z^2\big[\,\delta_{ij}\cosh^2{au\/2}+(\delta_{ij}-2k_ik_j)\sinh^2{au\/2}\,\big]\,\bigg\}\;,
\end{eqnarray}
where $u=\tau-\tau'$, the unit vector $k^\mu=(0,1,0,0)$, and it points along the
direction of acceleration. Therefore, from Eqs.~(\ref{cf}) and
(\ref{xf}), the boundary-dependent statistical functions of the
electric field can be expressed as
\begin{eqnarray}
(C_{ij}^F)_{bnd}(x(\tau),x(\tau'))&=&-{a^4\/32\pi^2}\biggl({1\/[\,\sinh^2{a\/2}(u-i\varepsilon)-a^2z^2]^3
}+{1\/[\,\sinh^2{a\/2}(u+i\varepsilon)-a^2z^2\,]^3}\biggr)\nonumber\\
&&\times \bigg\{\big[\,\delta_{ij}-2n_in_j+2az(n_ik_j+k_in_j)
\,\big]\sinh^2{au\/2}\nonumber\\
&& \quad \quad +
\,a^2z^2\big[\,\delta_{ij}\cosh^2{au\/2}+(\delta_{ij}-2k_ik_j)\sinh^2{au\/2}\,\big]\,\bigg\}\;,\label{acf}
\end{eqnarray}and
\begin{eqnarray}
(\chi_{ij}^F)_{bnd}(x(\tau),x(\tau'))&=&-{a^4\/32\pi^2}\biggl({1\/[\,\sinh^2{a\/2}(u-i\varepsilon)-a^2z^2]^3
}-{1\/[\,\sinh^2{a\/2}(u+i\varepsilon)-a^2z^2\,]^3}\biggr)\nonumber\\
&&\times \bigg\{\big[\,\delta_{ij}-2n_in_j+2az(n_ik_j+k_in_j)
\,\big]\sinh^2{au\/2}\nonumber\\
&& \quad \quad +
\,a^2z^2\big[\,\delta_{ij}\cosh^2{au\/2}+(\delta_{ij}-2k_ik_j)\sinh^2{au\/2}\,\big]\,\bigg\}\;.\label{axf}
\end{eqnarray}
Here one can see that for these statistical functions of the
electric field,  the diagonal components (the $xx$, $yy$ and $zz$
components), and the off-diagonal $xz$ component are nonzero. With
the statistical functions given above, we will use the residue
theorem to calculate separately the contributions of vacuum
fluctuations and radiation reaction to boundary-dependent energy
shifts. To be generic, we assume that the accelerated atom is
arbitrarily polarized. Thus we need to consider all the nonzero
statistical functions in our calculations.

\begin{figure}[htbp]
\centering \subfigure[]{
\includegraphics[width=3.0in]{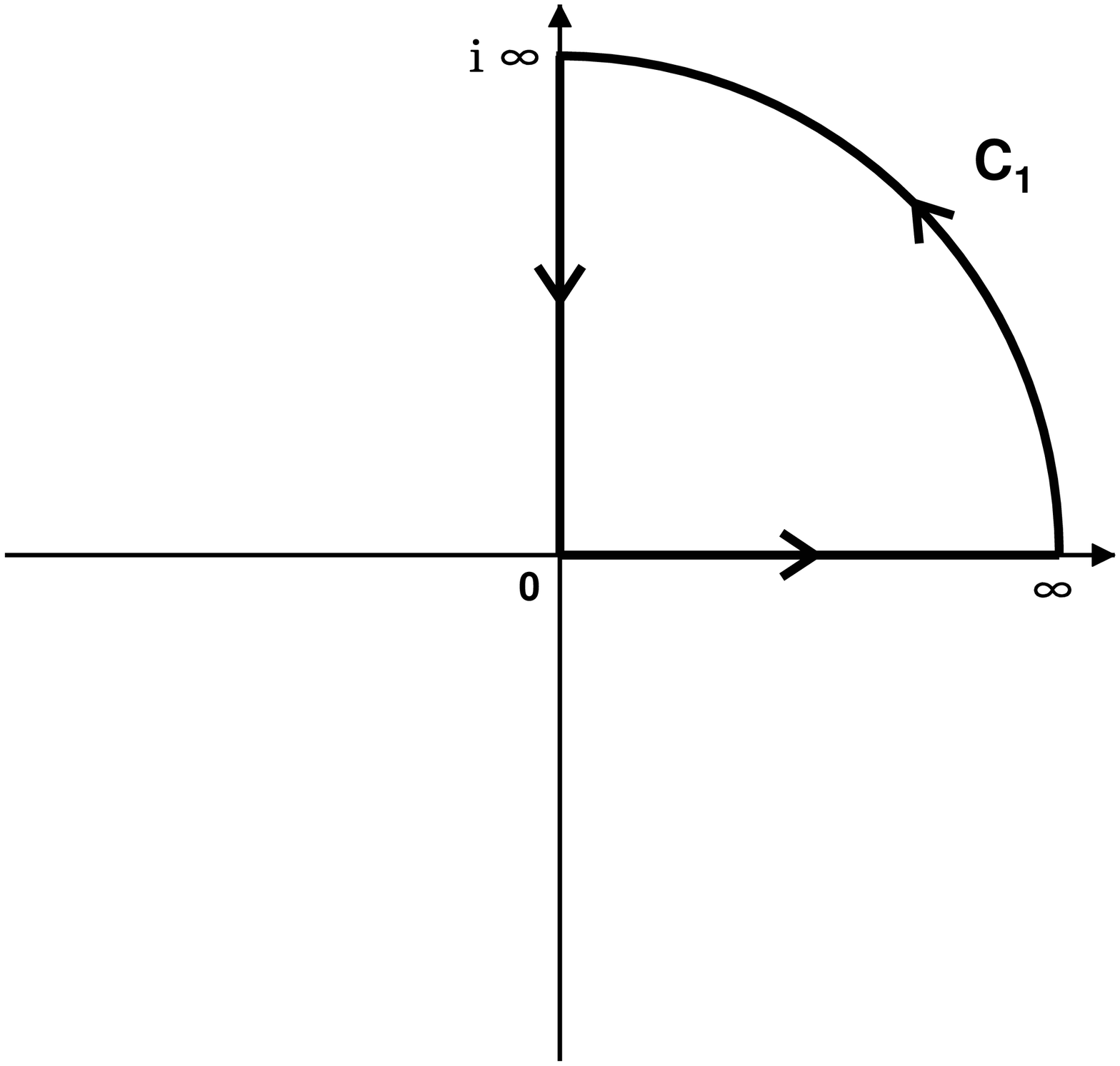}}
\subfigure[]{
\includegraphics[width=3.0in]{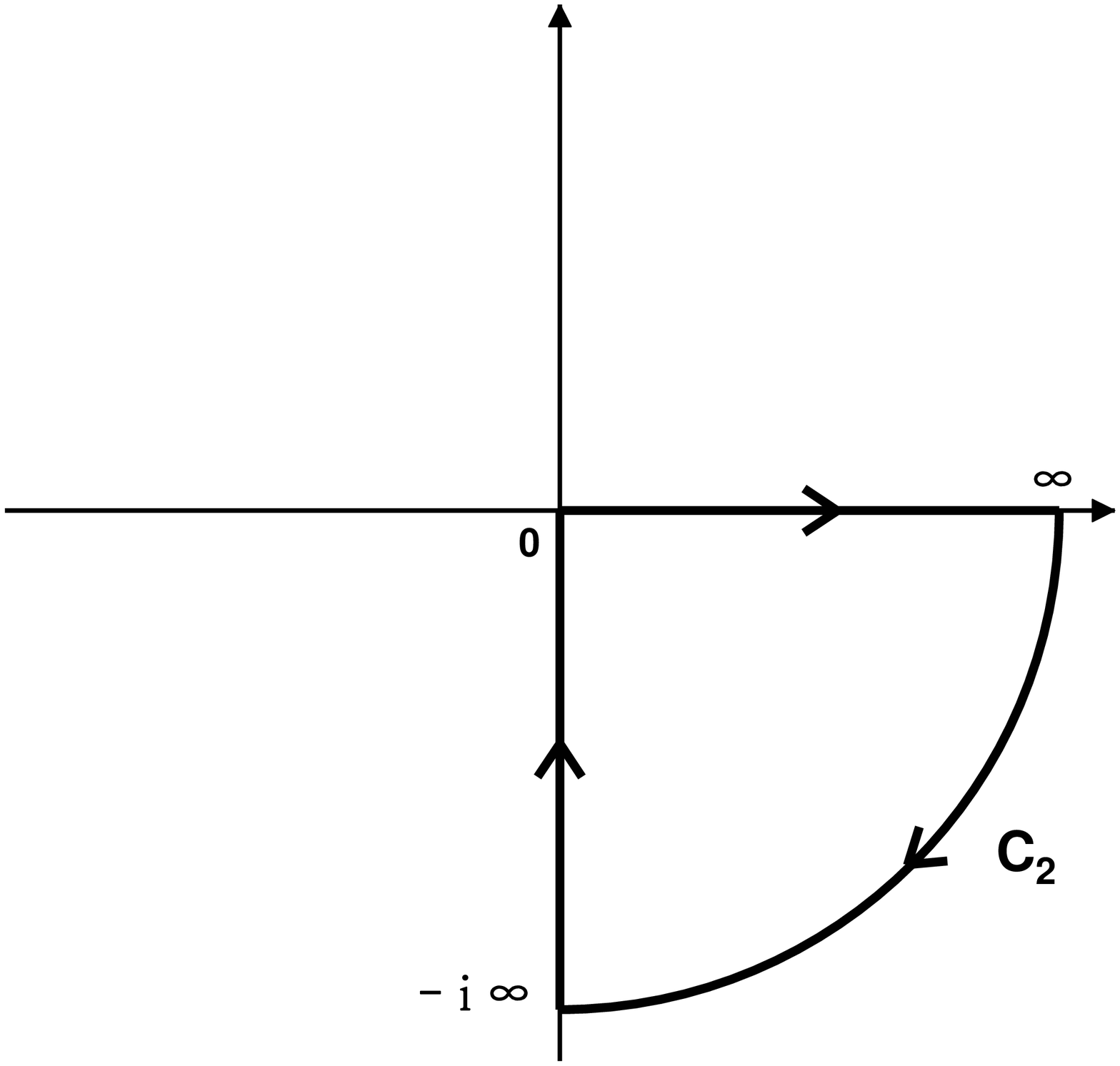}}
\caption{Integration path of  Eqs.~(\ref{cpvfxx}) and
(\ref{cprrxx}).}\label{fig1}
\end{figure}

Now we take the $xx$ component for an example to show how the
calculations are to be carried out. Substituting the statistical
functions (\ref{ca}), (\ref{xa}), (\ref{acf}) and (\ref{axf}) into
the general formulas (\ref{Ecpvf}) and (\ref{Ecprr}) and letting
$i=j=x$, we can obtain
\begin{eqnarray}
(\delta E_b)^{bnd}_{vf,xx}&=&{ia^4\/64\pi^2}\sum_d|\langle
b|\mu_x(0)|d\rangle|^2 \int_0^\infty
du(e^{i\omega_{bd}u}-e^{-i\omega_{bd}u})\nonumber\\&&\times\bigg({\sinh^2{au\/2}+a^2z^2\/[\sinh^2{a(u-i\varepsilon)\/2}-a^2z^2]^3}+
{\sinh^2{au\/2}+a^2z^2\/[\sinh^2{a(u+i\varepsilon)\/2}-a^2z^2]^3}\bigg)\;,\label{cpvfxx}
\end{eqnarray}
and
\begin{eqnarray}
(\delta E_b)^{bnd}_{rr,xx}&=&{ia^4\/64\pi^2}\sum_d|\langle
b|\mu_x(0)|d\rangle|^2 \int_0^\infty
du(e^{i\omega_{bd}u}+e^{-i\omega_{bd}u})\nonumber\\&&\times\bigg({\sinh^2{au\/2}+a^2z^2\/[\sinh^2{a(u-i\varepsilon)\/2}-a^2z^2]^3}-
{\sinh^2{au\/2}+a^2z^2\/[\sinh^2{a(u+i\varepsilon)\/2}-a^2z^2]^3}\bigg)\;,\label{cprrxx}
\end{eqnarray}
where we have extended the range of integration to infinity for
sufficiently long times $\tau-\tau_0$. For convenience, we take the
atom to be in its ground state. In order to evaluate the integral in
Eqs.~(\ref{cpvfxx}) and (\ref{cprrxx}), we use the residue theorem
and consider the contour integral along path $C_1$ and $C_2$ in
Fig.~\ref{fig1}.
With a definition of the atomic static scalar polarizability
\begin{eqnarray}
\alpha_0=\sum_{j}\alpha_j=\sum_{j~d}{2|\langle
b|\mu_j(0)|d\rangle|^2\/3\omega_{0}}\;,\label{alpha}
\end{eqnarray}
 the contribution of vacuum fluctuations to the position-dependent
energy shift of the ground state is calculated to be,
\begin{eqnarray}
(\delta
E_-)^{bnd}_{vf,xx}=-{3\omega_0\alpha_x\/128\pi}\bigg[\bigg(1+{2\/e^{2\pi\omega_0/a}-1}\bigg)f_{xx}(\omega_0,z,a)
-g_{xx}(\omega_0,z,a)\bigg] \label{Ecpvfxx2}\;,
\end{eqnarray}
and that of radiation reaction
\begin{eqnarray}
(\delta
E_-)^{bnd}_{rr,xx}={3\omega_0\alpha_x\/128\pi}f_{xx}(\omega_0,z,a)\;,
\end{eqnarray}
where we have defined
\begin{eqnarray}
f_{xx}(\omega_0,z,a)&=&{4z^2\omega_0^2(1+a^2z^2)-4a^4z^4-2a^2z^2-1\/z^3(1+a^2z^2)^{5/2}}\cos\bigg({2\omega_0\sinh^{-1}(az)\/a}\bigg)
\nonumber\\&&-{2\omega_0(1+4a^2z^2)\/z^2(1+a^2z^2)^2}\sin\bigg({2\omega_0\sinh^{-1}(az)\/a}\bigg)\;,\label{fxx}
\end{eqnarray}
and
\begin{eqnarray}
g_{xx}(\omega_0,z,a)={4a^4\/\pi}\int_0^\infty d
u{\sin^2{au\/2}-a^2z^2\/(\sin^2{au\/2}+a^2z^2)^3}e^{-\omega_0u}\;.\label{gxx}
\end{eqnarray}

Other nonzero components can be computed similarly. In summary, for
the case of an accelerated groud-state atom in the vicinity of an
infinite conducting plane, the contribution of vacuum fluctuations
to the position-dependent energy shift is given by
\begin{eqnarray}
(\delta E_-)^{bnd}_{vf}=-{3\omega_0\sqrt{\alpha_i\alpha_j}\/128\pi}
\bigg[\bigg(1+{2\/e^{2\pi\omega_0/a}-1}\bigg)f_{ij}(\omega_0,z,a)-g_{ij}(\omega_0,z,a)\bigg]\;,
\label{Ecpvfg}
\end{eqnarray}
while the contribution of radiation reaction is
\begin{eqnarray}
(\delta
E_-)^{bnd}_{rr}={3\omega_0\sqrt{\alpha_i\alpha_j}\/128\pi}f_{ij}(\omega_0,z,a)\;.\label{Ecprrge}
\end{eqnarray}
Here summation over repeated indices, $i$, $j$, is implied. The
nonzero functions $f_{xx}(\omega_0,z,a)$ and $g_{xx}(\omega_0,z,a)$
are given by Eqs.~(\ref{fxx}) and (\ref{gxx}), and the others are
\begin{eqnarray}
f_{yy}(\omega_0,z,a)&=&{4z^2\omega_0^2(1+a^2z^2)-1\/z^3(1+a^2z^2)^{3/2}}\cos\bigg({2\omega_0\sinh^{-1}(az)\/a}\bigg)
\nonumber\\&&-{2\omega_0(1+2a^2z^2)\/z^2(1+a^2z^2)}\sin\bigg({2\omega_0\sinh^{-1}(az)\/a}\bigg)\;,
\end{eqnarray}
\begin{eqnarray}
f_{zz}(\omega_0,z,a)&=&-{2+a^2z^2[5-4z^2\omega_0^2(1+a^2z^2)]\/z^3(1+a^2z^2)^{5/2}}\cos\bigg({2\omega_0\sinh^{-1}(az)\/a}\bigg)
\nonumber\\&&-{2\omega_0(2+a^2z^2+2a^4z^4)\/z^2(1+a^2z^2)^2}\sin\bigg({2\omega_0\sinh^{-1}(az)\/a}\bigg)\;,
\end{eqnarray}
\begin{eqnarray}
f_{xz}(\omega_0,z,a)&=&{a[1+4a^2z^2+4z^2\omega_0^2(1+a^2z^2)]\/z^2(1+a^2z^2)^{5/2}}\cos\bigg({2\omega_0\sinh^{-1}(az)\/a}\bigg)
\nonumber\\&&+{2a\omega_0(1-2a^2z^2)\/z(1+a^2z^2)^2}\sin\bigg({2\omega_0\sinh^{-1}(az)\/a}\bigg)\;,
\end{eqnarray}
\begin{eqnarray}
g_{yy}(\omega_0,z,a)={4a^4\/\pi}\int_0^\infty d
u{\sin^2{au\/2}-a^2z^2\cos(au)\/(\sin^2{au\/2}+a^2z^2)^3}e^{-\omega_0u}\;,
\end{eqnarray}
\begin{eqnarray}
g_{zz}(\omega_0,z,a)=-{4a^4\/\pi}\int_0^\infty d
u{\sin^2{au\/2}+a^2z^2\cos(au)\/(\sin^2{au\/2}+a^2z^2)^3}e^{-\omega_0u}\;,
\end{eqnarray}
\begin{eqnarray}
g_{xz}(\omega_0,z,a)={8a^5z\/\pi}\int_0^\infty d
u{\sin^2{au\/2}\/(\sin^2{au\/2}+a^2z^2)^3}e^{-\omega_0u}\;.
\end{eqnarray}
One can see clearly that both vacuum fluctuations  and radiation
reaction depend on the atom's acceleration. This differs from the
thermal corrections to the energy shifts of a static atom
\cite{Zhu-Yu09}, where radiation reaction is independent of the
temperature. A remarkable feature worth noting is that
 the position-dependent
energy shift for an atom polarized in the $x-z$ plane gets an extra
contribution associated with the functions $f_{xz}$ and $g_{xz}$ as
compared with that of the static case \cite{Zhu-Yu09,MJH}. This is
in qualitative agreement with the result in Ref.~\cite{Rizzuto09},
and it opens up the possibility of observing the effect of
acceleration of the atomic energy level shifts using atoms with
anisotropic polarization as pointed out in Ref.~\cite{Rizzuto09}.
Adding up the contributions of vacuum fluctuations and the radiation
reaction, we can obtain the total position-dependent energy shift of
the ground state
\begin{eqnarray}
(\delta E_-)^{bnd}_{tot}=-{3\omega_0\sqrt{\alpha_i\alpha_j}\/128\pi}
\bigg[{2\/e^{2\pi\omega_0/a}-1}f_{ij}(\omega_0,z,a)-g_{ij}(\omega_0,z,a)\bigg]\;.
\label{Ecpg}
\end{eqnarray}
If the atom's acceleration equals to zero, we can recover the
results in Ref~\cite{MJH}, which give the energy level shifts of a
static atom near a conducting plane in vacuum. Now one can see that
our expression for the position-dependent energy shift is in a
simpler and almost closed form as compared to that given in
Ref.~\cite{Rizzuto09} where the energy shift is written as integral
of a very complicated function(refer to Eqs.(29) and (30) of
Ref.~\cite{Rizzuto09} ). So, our result seems to be easier to handle
in terms of both analytical and numerical analysis. In what follows,
we examine the behaviors of the energy shift in various
circumstances and compare our results with those in the case of a
static atom in a thermal bath at the the Unruh temperature related
to the acceleration \cite{Zhu-Yu09}.

\section{discussion}
Now we examine the behaviors of the position-dependent energy shift
in the limits of low acceleration (when the atom's acceleration is
much smaller than the transition frequency of the atom,
$a\ll\omega_0$) and high acceleration (when the atom's acceleration
is much larger than the transition frequency of the atom,
$a\gg\omega_0$), analogously with the low- and high-temperature
limits $T\ll\omega_0$ and $T\gg\omega_0$, in Ref.~\cite{Zhu-Yu09}.
With the method of numerical analysis, we also analyze the case when
$a\sim\omega_0$.

\subsection{Low-acceleration limit}

In the low-acceleration limit $a\ll\omega_0$, we can identify the
distance between the atom and the boundary into three different
regimes: the short distance, where the distance $z$ is so small that
$az\ll\omega_0 z\ll 1$; the intermediate distance, where $az\ll
1\ll\omega_0 z$; and the long distance, where the distance $z$ is so
large that $\omega_0 z\gg az\gg1$. For these three distinct regimes,
we will discuss the behavior of the position-dependent energy shift
of a ground state atom.

Let us first consider the short distance regime, where
$az\ll\omega_0 z\ll 1$.  In this case, the contributions of vacuum
fluctuations and radiation reaction to the energy shift of a ground
state can be expressed as
\begin{eqnarray}
(\delta
E_-)^{bnd}_{vf}\approx-{1\/4\pi}\bigg[-{3\omega_0^2\alpha_z\/4\pi
z^2} +{3a\omega_0^2\sqrt{\alpha_x\alpha_z}\/4\pi z}
-{\omega_0^2(a^2+\omega_0^2)\/\pi}\ln(2\omega_0z)(\alpha_x+\alpha_y-\alpha_z)\bigg]\;,\label{lowasmallzvf}
\end{eqnarray}
and
\begin{eqnarray}
(\delta
E_-)^{bnd}_{rr}\approx-{1\/4\pi}\bigg[{3\omega_0(\alpha_x+\alpha_y+2\alpha_z)\/32z^3}
-{3a^2\omega_0\/64z}\bigg(\alpha_x+3\alpha_y+32\omega_0^2z^2\alpha_z+{2\/az}\sqrt{\alpha_x\alpha_z}\bigg)
\bigg]\;.\label{lowasmallzrr}\nonumber\\
\end{eqnarray}
Here the contribution of radiation reaction is much larger than that
of vacuum fluctuations and plays the dominant role in the total
energy shift. Note that  the acceleration induced correction term
associated with the atomic polarization in the $z$ direction in
Eq.~(\ref{lowasmallzrr}) is much smaller than that in
Eq.~(\ref{lowasmallzvf}). So the total energy shift containing the
acceleration corrections can be written approximately as
\begin{eqnarray}
(\delta
E_-)^{bnd}_{tot}\approx{1\/4\pi}\!\!&\bigg[&\!\!-{3\omega_0(\alpha_x+\alpha_y+2\alpha_z)\/32z^3}
\nonumber\\&&\!\!+{3a^2\omega_0\/64z}\bigg(\alpha_x+3\alpha_y+{64\omega_0z\ln(2\omega_0z)\/3}\alpha_z
+{2\/az}\sqrt{\alpha_x\alpha_z}\bigg)\bigg]\;.\label{lowasmallz}
\end{eqnarray}
The first (leading) term on the righthand side of
Eq.~(\ref{lowasmallz}) is just the energy shift of a static atom
interacting with a vacuum electromagnetic field near an infinite
conducting plane \cite{MJH}. The total energy shift is always
negative, and both the leading term and the acceleration correction
terms depend on the direction of the atom's polarization. For an
 isotropically polarized atom, the off-diagonal $xz$ component
contributes the biggest acceleration correction term and the
position-dependent energy shift can be written as
\begin{eqnarray}
(\delta
E_-)^{bnd}_{tot}\approx-{1\/4\pi}\bigg({\omega_0\alpha_0\/8z^3}
-{\omega_0\alpha_0a\/32z^2}\bigg)\;,\label{lowasmallz2}
\end{eqnarray}
where the first (leading) term is just the result of an inertial
atom in the vacuum, and the acceleration correction term is
proportional to the atom's acceleration. These two terms are
opposite in sign. With the Unruh temperature $T_U=a/(2\pi)$,
Eq.~(\ref{lowasmallz2}) becomes
\begin{eqnarray}
(\delta
E_-)^{bnd}_{tot}\approx-{1\/4\pi}\bigg({\omega_0\alpha_0\/8z^3}
-{\pi\omega_0\alpha_0\/16z^2}T_U\bigg)\;,
\end{eqnarray}
which grows linearly with the Unruh temperature. Let us recall the
position-dependent energy shift of a static atom at low temperature
and in the short distance regime~\cite{Zhu-Yu09}, where the thermal
correction is dominated by the term proportional to $T^4$.
Therefore, in this case the acceleration effect is not equal to the
thermal effect, although the total energy shifts agree in the
leading order.

Then in the intermediate distance regime, where $az\ll 1\ll\omega_0
z$, we find, approximately, the contributions of vacuum fluctuations
and radiation reaction,
\begin{eqnarray}
(\delta
E_-)^{bnd}_{vf}\approx-{1\/4\pi}\!\!&\bigg\{&\!\!\!\bigg[{3\omega_0^3\/8z}\bigg(\alpha_x+\alpha_y-{1\/2z^2\omega_0^2}\alpha_z\bigg)
\nonumber\\&&\!\!\!-{3\omega_0^3a^2z\/16}\bigg(3\alpha_x+\alpha_y-2\alpha_z-{2\sqrt{\alpha_x\alpha_z}\/az}\bigg)\bigg]
\cos(2\omega_0z)\nonumber\\&&\!\!\!-\bigg[{3\omega_0^2(\alpha_x+\alpha_y+2\alpha_z)\/16z^2}
+{3\omega_0^2a^2\/16}\bigg(2\alpha_x+\alpha_y-3\alpha_z-{\sqrt{\alpha_x\alpha_z}\/az}\bigg)\bigg]\sin(2\omega_0z)
\nonumber\\&&\!\!\!+{3\/8\pi z^4}(\alpha_x+\alpha_y+\alpha_z)
+{3a^2\/8\pi\omega_0^2z^4}\bigg({2\alpha_x\/\omega_0^2z^2}-\alpha_y-\alpha_z-{\sqrt{\alpha_x\alpha_z}\/az}\bigg)\bigg\}\;,
\end{eqnarray}
and
\begin{eqnarray}
(\delta
E_-)^{bnd}_{rr}\approx{1\/4\pi}\!\!&\bigg\{&\!\!\!\bigg[{3\omega_0^3\/8z}\bigg(\alpha_x+\alpha_y-{1\/2z^2\omega_0^2}\alpha_z\bigg)
\nonumber\\&&\!\!\!-{3\omega_0^3a^2z\/16}\bigg(3\alpha_x+\alpha_y-2\alpha_z-{2\sqrt{\alpha_x\alpha_z}\/az}\bigg)\bigg]
\cos(2\omega_0z)\nonumber\\&&\!\!\!-\bigg[{3\omega_0^2(\alpha_x+\alpha_y+2\alpha_z)\/16z^2}
+{3\omega_0^2a^2\/16}\bigg(2\alpha_x+\alpha_y-3\alpha_z-{\sqrt{\alpha_x\alpha_z}\/az}\bigg)\bigg]\sin(2\omega_0z)
\bigg\}\;,\nonumber\\
\end{eqnarray}
where both  have terms which are oscillating functions of $z$.
However, when we add up these two contributions, the oscillating
components cancel out, and we find
\begin{eqnarray}
(\delta E_-)^{bnd}_{tot}\approx-{1\/4\pi}\bigg[{3\/8\pi
z^4}(\alpha_x+\alpha_y+\alpha_z)
+{3a^2\/8\pi\omega_0^2z^4}\bigg({2\alpha_x\/\omega_0^2z^2}-\alpha_y-\alpha_z-{\sqrt{\alpha_x\alpha_z}\/az}\bigg)\bigg]\;.\label{totlowaint}
\end{eqnarray}
Here the first term plays the dominant role in the
position-dependent energy shift, and it is agreement with that of a
static atom placed far from the boundary in the vacuum \cite{MJH}.
For an isotropically polarized atom, the acceleration correction
comes mainly from the off-diagonal $xz$ component under the
condition $az\ll 1\ll\omega_0 z$, and the position-dependent energy
shift becomes
\begin{eqnarray}
(\delta E_-)^{bnd}_{tot}\approx-{1\/4\pi}\bigg({3\alpha_0\/8\pi z^4}
-{\alpha_0a\/8\pi\omega_0^2z^5}\bigg)=-{1\/4\pi}\bigg({3\alpha_0\/8\pi
z^4} -{\alpha_0\/4\omega_0^2z^5}T_U\bigg)\;.\label{totlowaint2}
\end{eqnarray}
Here the position-dependent energy shift is still negative, and
grows linearly with the Unruh temperature, while in the case of a
static atom immersed in a thermal bath, the temperature correction
is proportional to $T^4$ in the intermediate distance regime for the
low temperature limit~\cite{Zhu-Yu09}.

Finally, it is time to examine the case of the atom in the long
distance regime ($\omega_0 z\gg az\gg1$). In order to compare the
accelerated case with the thermal case in this regime, let us first
recall the energy shift of an isotropically polarized atom in a
thermal bath in the long distance regime in the low temperature
limit~\cite{Zhu-Yu09},
\begin{eqnarray}
(\delta E_-)^{bnd}_{tot}\approx-{1\/4\pi}{\alpha_0T\/4z^3}\;.
\end{eqnarray}
If an accelerated atom were to behave the same as a static one
immersed in a thermal bath at the Unruh temperature $T_U=a/(2\pi)$,
we would expect that the energy shift of the accelerated atom should
be proportional to the atom's acceleration. But, in fact,  the
energy shift (\ref{Ecpg}) can also be approximated, in the
long-distance regime for the low-acceleration limit,  by
Eq.~(\ref{totlowaint2}), where the acceleration just induces a very
small correction, and this correction term, although linear in the
Unruh temperature,  is not the leading term as in the thermal case.
This shows that the behavior of an accelerated atom in the long
distance regime is completely different from that of the static one
immersed in a thermal bath at the Unruh temperature.

\subsection{High-acceleration limit}

Now let us turn our attention to the high-acceleration limit, where
we assume that the atom's acceleration is much larger than the
transition frequency of the atom, $a\gg\omega_0$. In this limit the
contribution of vacuum fluctuations to the energy shift of the
ground state, Eq.~(\ref{Ecpvfg}), can be written approximately as
\begin{eqnarray}
(\delta E_-)^{bnd}_{vf}=-{3\omega_0\sqrt{\alpha_i\alpha_j}\/128\pi}
\bigg({a\/\pi\omega_0}f_{ij}(\omega_0,z,a)-g_{ij}(\omega_0,z,a)\bigg)\;.
\end{eqnarray}
As in the low-acceleration limit,  we will analyze the behavior of
the energy-level shift in three different regimes. Let us first deal
with the case when the atom is so close to the boundary that
$\omega_0z\ll az\ll1$ (the short distance regime). It then follows
that
\begin{eqnarray}
(\delta E_-)^{bnd}_{vf}\approx{1\/4\pi}{3a\/32\pi
z^3}(\alpha_x+\alpha_y+2\alpha_z-az\sqrt{\alpha_x\alpha_z})\;,\label{highashortvf}
\end{eqnarray}
and
\begin{eqnarray}
(\delta
E_-)^{bnd}_{rr}\approx-{1\/4\pi}\bigg[{3\omega_0(\alpha_x+\alpha_y+2\alpha_z)\/
32z^3}-{3\omega_0a\sqrt{\alpha_x\alpha_z}\/32z^2}-{3\omega_0a^2(\alpha_x+3\alpha_y)\/64z}
-{45\omega_0a^4z\alpha_z\/128}\bigg]\;.\nonumber\\\label{highashortrr}
\end{eqnarray}
Note that the contribution of vacuum fluctuations comes mainly from
the acceleration correction terms, while  the acceleration just
gives very small corrections to that of the radiation reaction.
Under the high acceleration condition ($a\gg\omega_0$), the
contribution of vacuum fluctuations is much larger than that of
radiation reaction and plays the dominant role in the
boundary-dependent energy shift. For an isotropically polarized atom
,  the total energy shift becomes
\begin{eqnarray}
(\delta E_-)^{bnd}_{tot}\approx{1\/4\pi}\bigg({\alpha_0a\/8\pi
z^3}-{\alpha_0a^2\/32\pi z^2}\bigg)={1\/4\pi}\bigg({\alpha_0\/4
z^3}T_U-{\pi\alpha_0\/8 z^2}T_U^2\bigg) \;,\label{highashorttot}
\end{eqnarray}
where the first (leading) term is linear in the acceleration of the
atom and is the same as the corresponding result of an inertial atom
immersed in a thermal bath at the Unruh temperature \cite{Zhu-Yu09}.
But the subleading term that comes from the $xz$ component is absent
in the thermal case. So, in terms of the energy level shifts, the
accelerated behaves as if immersed in a thermal bath at the Unruh
temperature  in the leading order. This differs from the
corresponding result in the scalar field case in the high
acceleration limit, where the energy shift is independent of the
acceleration and differs completely  from a static atom immersed in
a thermal bath at the Unruh temperature \cite{wtzhou10}.

If we increase, in the high acceleration limit ($a\gg\omega_0$), the distance $z$ between the atom and the
conducting boundary, such that $az\gg1$, the contributions of vacuum
fluctuations and radiation reaction can be approximated as
\begin{eqnarray}
(\delta E_-)^{bnd}_{vf}&\approx&-{1\/4\pi}\bigg\{-{3\alpha_x\/8\pi z^4}\bigg[\bigg(1-{\omega_0^2\/a^2}\bigg)\cos{2\omega_0\/a}\ln(2az)
+{2\omega_0\/a}\sin{2\omega_0\/a}\ln(2az)-1\bigg]
\nonumber\\&&+{3\omega_0^2\alpha_y\/8\pi z^2}\bigg[\bigg(1-{1\/4\omega_0^2a^2z^4}\bigg)
\cos{2\omega_0\/a}\ln(2az)-{a\/\omega_0}\sin{2\omega_0\/a}\ln(2az)+{1\/a^2z^2}+{1\/4\omega_0^2a^2z^4}\bigg]
\nonumber\\&&+{3\omega_0^2\alpha_z\/8\pi z^2}\bigg[\bigg(1-{5\/4\omega_0^2a^2z^4}\bigg)
\cos{2\omega_0\/a}\ln(2az)-{a\/\omega_0}\sin{2\omega_0\/a}\ln(2az)+{1\/a^2z^2}+{5\/4\omega_0^2a^2z^4}\bigg]
\nonumber\\&&+{3\omega_0^2\sqrt{\alpha_x\alpha_z}\/8\pi a z^3}\bigg[\cos{2\omega_0\/a}\ln(2az)
-{a\/\omega_0}\sin{2\omega_0\/a}\ln(2az)-{1\/\omega_0^2z^2}\bigg]\bigg\}
\;,\label{highaintvf0}
\end{eqnarray}
and
\begin{eqnarray}
(\delta E_-)^{bnd}_{rr}&\approx&{1\/4\pi}\bigg\{-{3\omega_0\alpha_x\/8a z^4}\bigg[\bigg(1-{\omega_0^2\/a^2}\bigg)\cos{2\omega_0\/a}\ln(2az)
+{2\omega_0\/a}\sin{2\omega_0\/a}\ln(2az)\bigg]
\nonumber\\&&+{3\omega_0^3\alpha_y\/8a z^2}\bigg[\bigg(1-{1\/4\omega_0^2a^2z^4}\bigg)
\cos{2\omega_0\/a}\ln(2az)-{a\/\omega_0}\sin{2\omega_0\/a}\ln(2az)\bigg]
\nonumber\\&&+{3\omega_0^3\alpha_z\/8a z^2}\bigg[\bigg(1-{5\/4\omega_0^2a^2z^4}\bigg)
\cos{2\omega_0\/a}\ln(2az)-{a\/\omega_0}\sin{2\omega_0\/a}\ln(2az)\bigg]
\nonumber\\&&+{3\omega_0^3\sqrt{\alpha_x\alpha_z}\/8 a^2 z^3}\bigg[\cos{2\omega_0\/a}\ln(2az)
-{a\/\omega_0}\sin{2\omega_0\/a}\ln(2az)\bigg]\bigg\}\;,\label{highaintrr0}
\end{eqnarray}
where the approximation of $\sinh^{-1}(az)\sim\ln(2az)$ for $az\gg1$
is used. Here both the contributions of vacuum fluctuations and
radiation reaction contain terms that are oscillating functions of
the atom's acceleration and the distance $z$, and the amplitudes of
the oscillating functions in Eq.~(\ref{highaintrr0}) are much
smaller than those in Eq.~(\ref{highaintvf0}). Therefore, the
energy-level shift of the ground-state atom exhibits an oscillatory
behavior, which is typical of the energy-level shift of an excited
state. This can be understood as a result of the fact that the
ground-state atoms have a nonvanshing possibility to absorb thermal
photons to transit to the excited states in the high-temperature
limit, and it actually brings in an interesting issue of the thermal
average of energy-level shifts of an atom in thermal equilibrium.
Let us note that this issue, which we leave for future research, can
be addressed using our results in the present paper and the
spontaneous excitation rate found in Ref.~\cite{Yu-Zhu06} in a
similar way as what has been done in the case of a static atom in a
thermal bath~\cite{Zhu-Yu09}.

 If we further assume that
${2\omega_0\/a}\ln(2az)\ll1$, which is easily satisfied in the
limits $a\gg\omega_0$ and $az\gg1$, since the logarithm is a very
slowly varying function, the above results can be simplified to
\begin{eqnarray}
(\delta E_-)^{bnd}_{vf}\approx-{1\/4\pi}&\bigg\{&{3\omega_0^2\/8\pi
a^2z^4}\bigg[1-4\ln(2az)+2(\ln(2az))^2\bigg]\alpha_x\nonumber\\&&+{3\omega_0^2\/8\pi
z^2}[1-2\ln(2az)]\bigg(\alpha_y+\alpha_z+{1\/az}\sqrt{\alpha_x\alpha_z}\bigg)\bigg\}\;,\label{highaintvf}
\end{eqnarray}
and
\begin{eqnarray}
(\delta E_-)^{bnd}_{rr}\approx{1\/4\pi}&\bigg\{&-{3\omega_0\/8a
z^4}\alpha_x +{3\omega_0^3\/8a
z^2}[1-2\ln(2az)]\bigg(\alpha_y+\alpha_z+{1\/az}\sqrt{\alpha_x\alpha_z}\bigg)\nonumber\\&&
-
{3\omega_0\/32a^3z^6}(\alpha_y+5\alpha_z-4az\sqrt{\alpha_x\alpha_z})\bigg\}\;.\label{highaintrr}
\end{eqnarray}
 If the atom is
polarized only in the $x$-direction, the contribution of radiation
reaction is much larger than that of vacuum fluctuations. However,
if the atom's polarization is along the $y$ or $z$ direction, the
ratio of vacuum fluctuations part to radiation reaction part is
determined by the magnitude of the quantity $\omega_0z$. If
$\omega_0z\ll1$ (the intermediate distance regime), this ratio is
indeterminate with an indeterminate quantity $\omega_0a^3z^4$. Let
us note that this disparity between the $x$ and $y$ components does
not exist in the thermal case, and this is not surprising since  the
atom is accelerating in the $x$ direction anyway. If the atom is so
far from the boundary that $\omega_0z\gg1$ (the long distance
regime), the contributions of vacuum fluctuations associated with
the atomic polarization in the $y$  and $z$ directions are much
larger than that of radiation reaction, and the total energy shift
can be written as
\begin{eqnarray}
(\delta E_-)^{bnd}_{tot}\approx-{1\/4\pi}\bigg\{{3\omega_0\/8a
z^4}\alpha_x+{3\omega_0^2\/8\pi
z^2}[1-2\ln(2az)]\bigg(\alpha_y+\alpha_z+{1\/az}\sqrt{\alpha_x\alpha_z}\bigg)\bigg\}\;,\label{highainttot}
\end{eqnarray}
which is negative if the atom is polarized in the direction along
the atom's acceleration ($x$ direction) and positive if the atom is
polarized in the $y$  or $z$ direction. Written in terms of  the
Unruh temperature, the total energy shift becomes
\begin{eqnarray}
(\delta E_-)^{bnd}_{tot}\approx-{1\/4\pi}\bigg\{{3\omega_0\/16\pi
z^4T_U}\alpha_x+{3\omega_0^2\/8\pi z^2}[1-2\ln(4\pi
zT_U)]\bigg(\alpha_y+\alpha_z+{1\/2\pi
zT_U}\sqrt{\alpha_x\alpha_z}\bigg)\bigg\}\;,
\end{eqnarray}
which differs from the corresponding result of a static atom
immersed in a thermal bath at the Unruh temperature \cite{Zhu-Yu09}.

\subsection{The case of $a\sim\omega_0$}

Having discussed both the cases of low- ($a\ll\omega_0$) and
high-acceleration ($a\gg\omega_0$) limits, we now  consider the case
when $a\sim\omega_0$, since this is probably the typical
acceleration necessary to observe the Unruh effect. Although we will
mostly resort to numerical method for analysis, using our result
which is simpler than that of Ref.~\cite{Rizzuto09}, we are able to
obtain approximate analytical expressions in some special cases.

Let us  first examine what happens when the atom is  very close to
the boundary ($z\omega_0\ll1$). Now one can show that
\begin{eqnarray}
(\delta
E_-)^{bnd}_{tot}\approx-{3\omega_0\/128\pi}\bigg[{1\/z^3}(\alpha_x+\alpha_y+2\alpha_z)-{a\/z^2}\sqrt{\alpha_x\alpha_z}\bigg]\;,\label{z0}
\end{eqnarray}
where the leading term is coincident with the energy shift of a
static atom interacting with a vacuum electromagnetic field near an
infinite conducting plane \cite{MJH}. Here,  the acceleration
correction term comes from the contribution of radiation reaction
which is modified by the presence of the acceleration in sharp
contrast to the case of a static atom in a thermal bath and it is
also a result of  the off-diagonal $xz$ component of $f$ functions
which is absent in the thermal case. It is this off-diagonal term
which is unique to the case of an accelerated atom that makes the
energy shift smaller than that of a static one.

If the atom is far from the boundary ($z\omega_0\gg1$), then one
finds
\begin{eqnarray}
(\delta
E_-)^{bnd}_{tot}\approx-{3\omega_0\/128\pi}&\bigg\{&{2\/e^{2\pi}-1}\bigg[\bigg({2\alpha_x\/\omega_0^3z^6}
+{4\omega_0(\alpha_y+\alpha_z)\/z^2}+{4\sqrt{\alpha_x\alpha_z}\/z^3}\bigg)\cos(2\sinh^{-1}(az))
\nonumber\\&&-\bigg({8\alpha_x\/\omega_0z^4}+{4\omega_0(\alpha_y+\alpha_z)\/z^2}+{4\sqrt{\alpha_x\alpha_z}\/z^3}\bigg)\sin(2\sinh^{-1}(az))\bigg]
\nonumber\\&&+{2\/\pi\omega_0z^4}\bigg(2\alpha_x+\alpha_y+\alpha_z-{\sqrt{\alpha_x\alpha_z}\/\omega_0z}\bigg)\bigg\}\;,\label{zl}
\end{eqnarray}
which is an oscillating function of $z$.  The oscillatory behavior
here, which is reminiscent of the energy0level shift of a static
excited state,  can again be attributed to the nonvanishing
spontaneous excitation rate of a ground state when $a\gtrsim
\omega_0$.  It is to be noted that if the atom is polarized in the
$y$ and $z$ directions, the amplitude is proportional to $1/z^2$.
Recalling the case of a static atom where the energy shift is
proportional to $1/z^4$ when $z\omega_0\gg1 $\cite{MJH}, we see that
the energy shift of an accelerated atom with anisotropic
polarizability may be much larger than that of the static one when
the atom is far from the boundary. This may open up a possibility of
an indirect detection of the Unruh effect through the measurements
related to the energy-level shifts of accelerating atoms. It is
interesting to note that
a similar result has been recently obtained for the energy shift
between two accelerating atoms in Ref.~\cite{Marino}.

\begin{figure}[htbp]
\centering
\includegraphics[width=5.0in]{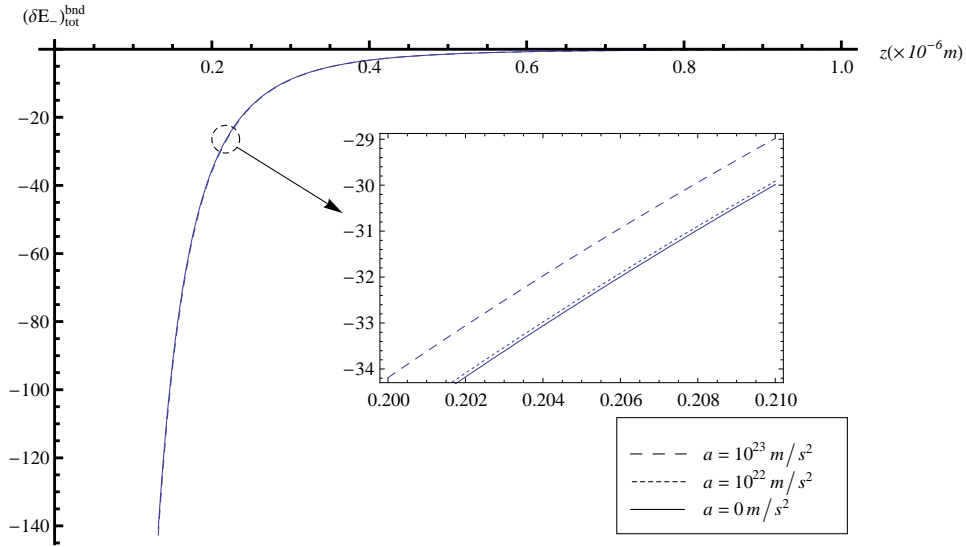}
\caption{Energy shifts as a function of the distance $z$ and for
different values of atomic acceleration. Here the typical value of
the atomic transition frequency, $\omega_0=10^{15}$\,s$^{-1}$, is
used and the energy shifts are in the units of
$\alpha_0/(128\pi\varepsilon_0)$. $\varepsilon_0$ is the vacuum
dielectric constant. The dashed, dotted and solid lines represent
the energy shifts for $a=10^{23}$\,m/s$^2$, $a=10^{22}$\,m/s$^2$ and
$a=0$\,m/s$^2$ respectively. For the distance $z$ on the range of
$0\sim 1\times10^{-6}$m, we can distinguish, by eye, the three lines
of different acceleration  in the figure from each other. But on an
extremely small range, one can see the discrepancy
clearly.}\label{fig8}
\end{figure}

Taking $\omega_0$ to be the typical transition frequency of a
hydrogen atom, i.e., $\omega_0\sim 10^{15}$\,s$^{-1}$.  we have
plotted, in Fig.~\ref{fig8},  the energy shift as a function of the
distance $z$ on the order of microns for a ground-state atom with
isotropic polarizability and for different values of atomic
acceleration.

A comparison between the different curves in Fig.~\ref{fig8} shows
that the effects of acceleration on the energy shift become
appreciable for accelerations on the order of $10^{23}$m/s$^2$, and
with the increase of the acceleration the absolute values of the
energy shift decrease. This is contrary to the conclusion drawn from
Fig.~$1$ in Ref.~\cite{Rizzuto09}, where the effect of acceleration
is found to make the energy shift larger than that of the atom at
rest and the absolute values of the energy shift increase with the
the increase of the acceleration~\footnote{Note the SI units are
adopted in all the figures of this paper, as opposed to  the CGS
units used in Ref.~\cite{Rizzuto09}.}.

\begin{figure}[htbp]
\centering
\includegraphics[width=5.0in]{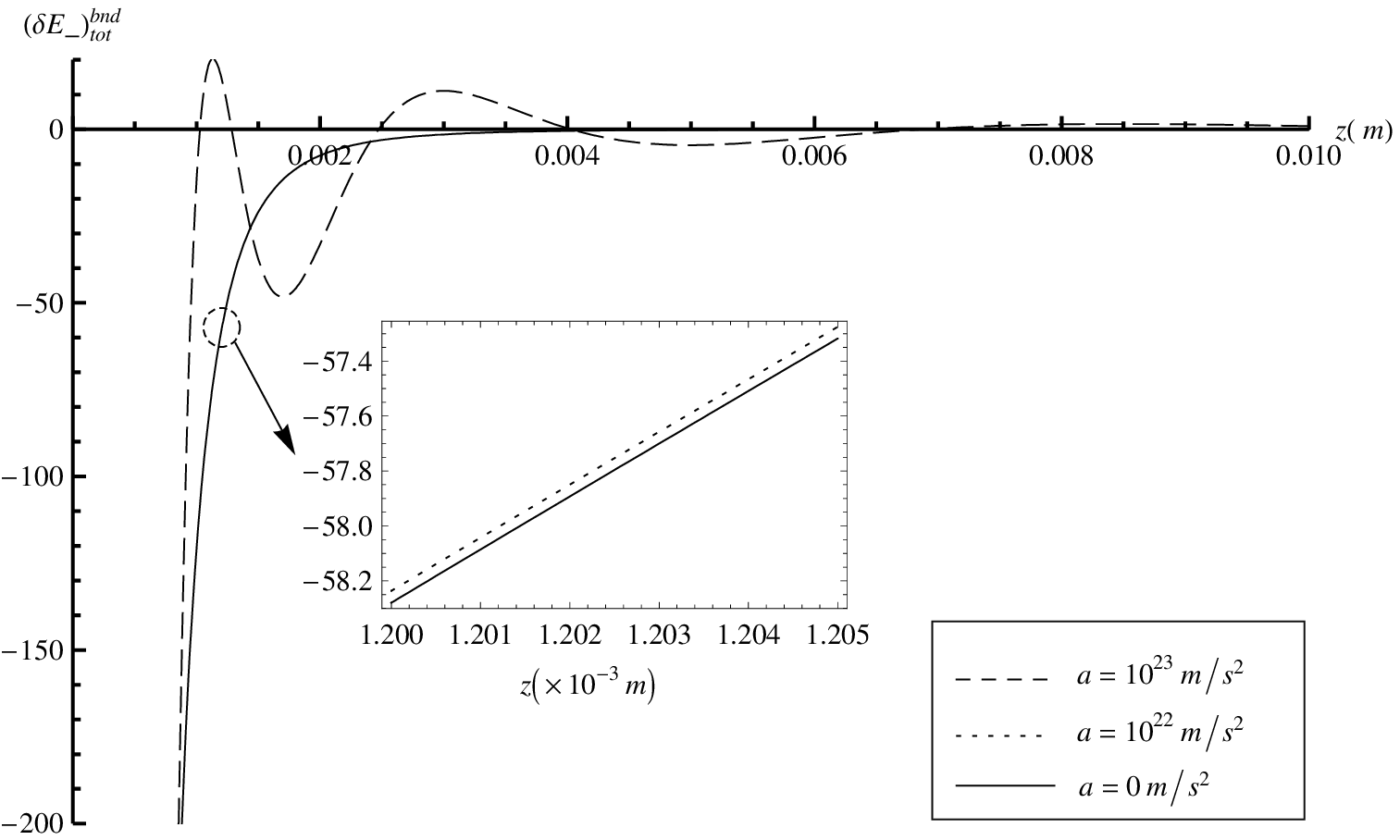}
\caption{Energy shifts as a function of the distance $z$ and for
different values of atomic acceleration. Here the typical value of
the atomic transition frequency, $\omega_0=10^{15}$\,s$^{-1}$ is
used and the energy shifts are in the units of
$\alpha_0/(128\pi\varepsilon_0\omega_0)$. Note that for the distance
$z$ on the range of $0\sim 0.01$m, we can not distinguish by eye the
lines for $a=10^{22}$\,m/s$^2$ and $a=0$\,m/s$^2$. But on an
extremely small range, we can see the discrepancy
clearly.}\label{fig2}
\end{figure}
 In Fig.~\ref{fig2}, we have plotted, on a larger distance scale, the energy shift
for a ground-state atom with isotropic polarizability and for
different values of atomic acceleration. From the figure, one can
see that when $a\sim 10^{22}$\,m/s$^2$, which is one order of
magnitude less than $\omega_0$, the effect of acceleration makes the
energy shift smaller than that of an atom at rest. But the
discrepancy between these two energy shifts is very small. This
numerical result is in agreement with that in the case of low
acceleration limit analytically discussed in the previous section,
where we find that the acceleration correction is small and opposite
in sign compared with the energy shift of a static atom. After this
consistency check, we now move to the more interesting case, i.e.,
when $a\sim\omega_0$. Then the corresponding acceleration $a\sim
10^{23}$\,m/s$^2$.  Now the Figure shows obvious oscillations of the
energy shift when distance $z\gtrsim 10^{-3}$m.
For $z\gtrsim10^{-3}$m, $z\omega_0\gg1 $ is satisfied and therefore
the energy shift can be approximated by the analytical expressions
in Eq.~(\ref{zl}). This oscillation gives rise to a clear difference
between the energy shift of an accelerated atom and that of a static
one. So, on a theoretical front, the effect of the atomic
acceleration on the energy shift now becomes appreciable. However,
it should be noted that, on an experimental front, a distance of $
\sim 10^{-3}$m appears to be unrealistic for  experimental
techniques, since all actual measurements of the atom-wall force
involve much shorter distances.

\begin{figure}[htbp]
\centering
\includegraphics[width=5.0in]{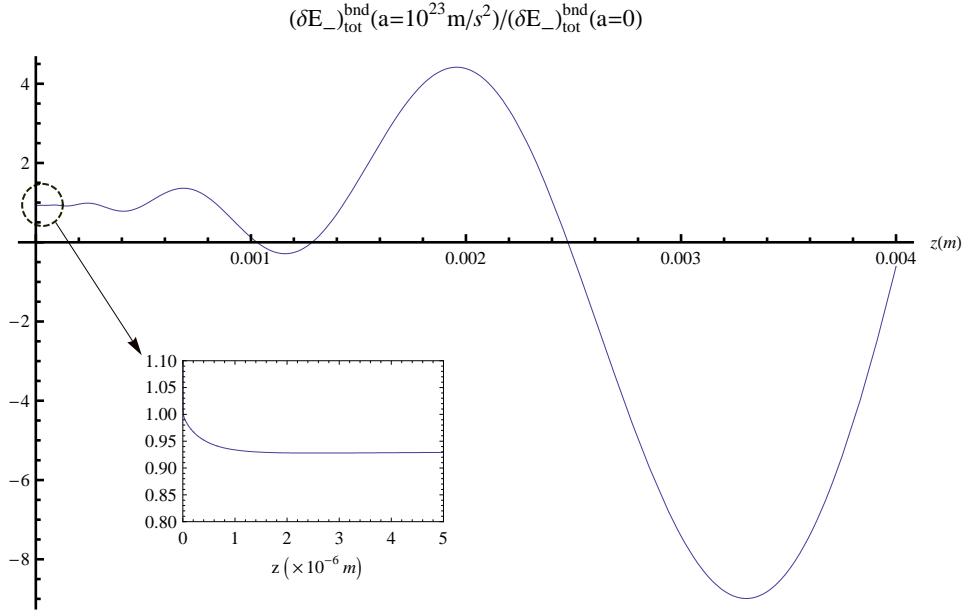}
\caption{The ratio between the energy shift of an accelerated atom
in front of an infinite conducting plane and that of a static one
for acceleration $a=10^{23}$\,m/s$^2$.}\label{fig3}
\end{figure}

In order to show the effect of the acceleration more clearly, let us
look at the ratio between the energy shift of an accelerated atom in
front of an infinite conducting plane and that of a static one
$(\delta E_-)^{bnd}_{tot}(a\neq0)/(\delta E_-)^{bnd}_{tot}(a=0)$ for
the typical transition frequency of a hydrogen atom,
$\omega_0\sim10^{15}$\,s$^{-1}$, and the corresponding acceleration
$a\sim10^{23}$\,m/s$^2$. We  plot the ratio as a function of the
distance $z$  in Fig.~\ref{fig3}, where an isotropic polarizability
of the atom is assumed.
The sub-figure shows that when the atom is close to the boundary,
i.e., when $z\ll10^{-6}$m, this ratio approaches to $1$, and the
energy shift of an accelerated atom approaches to that of a static
one. For example,  the ratio is  $\sim 0.997$ when $z\sim 10^{-8}$m.
With the increase of the distance $z$, we can see clearly the
oscillating behaviors of this ratio and the increasing amplitude of
the oscillation, although the energy shift of an accelerated atom
alone has an oscillating decay. It is interesting to note that the
absolute value of this ratio is much larger than $1$ for some values
of the distance $z$, and thus the effect of acceleration on the
position-dependent energy shift will be significant, whereas  there
also exist some values of the distance $z$ where the ratio equals
$1$, and, at these distances, the correction of acceleration
vanishes. Note that the value of this ratio is smaller than $1$ for
the distances at the order of $10^{-6}$m. In other words, the
acceleration makes the position-dependent energy shift smaller. This
 feature  differs from that obtained in
Ref.~\cite{Rizzuto09}, where  the energy shift of the accelerated
atom ($a=10^{23}$m/s$^2$) is found to be much larger than that of a
static one at the same distance scale(refer to Fig.~$2$ in
Ref.~\cite{Rizzuto09} and keep in mind that what we actually plot
here is the reciprocal of what is plotted there). At the same time,
our analysis also reveals an oscillatory behavior of the ratio on a
larger distance scale.

\begin{figure}[htbp] \centering
\includegraphics[width=5.0in]{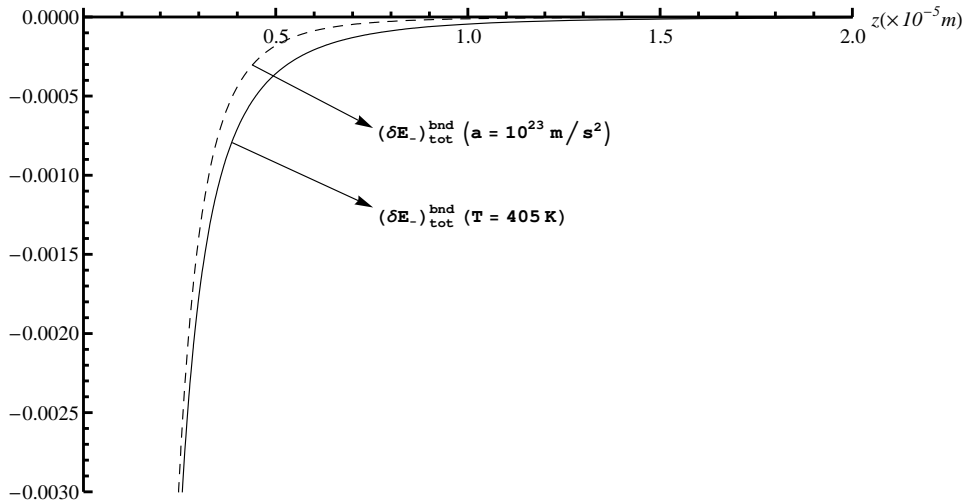}
\caption{Energy shifts of both an accelerated atom and a static one
immersed in a thermal bath at the corresponding Unruh temperature.
Here the typical value of the atomic transition  frequency,
$\omega_0=10^{15}$\,s$^{-1}$ is used and the energy shifts are in
the units of $\alpha_0/(128\pi\varepsilon_0)$.}\label{fig4}
\end{figure}

\begin{figure}[htbp]
\centering \subfigure[]{
\includegraphics[width=3.0in]{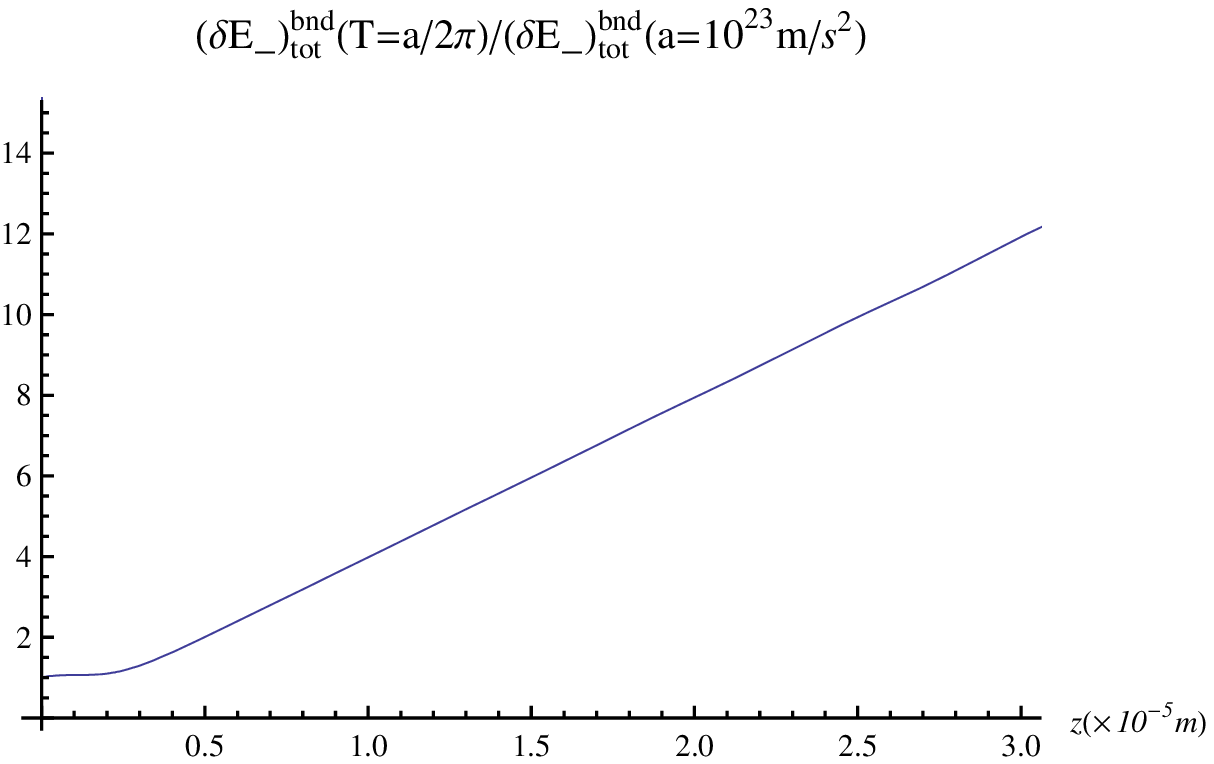}}
\subfigure[]{
\includegraphics[width=3.0in]{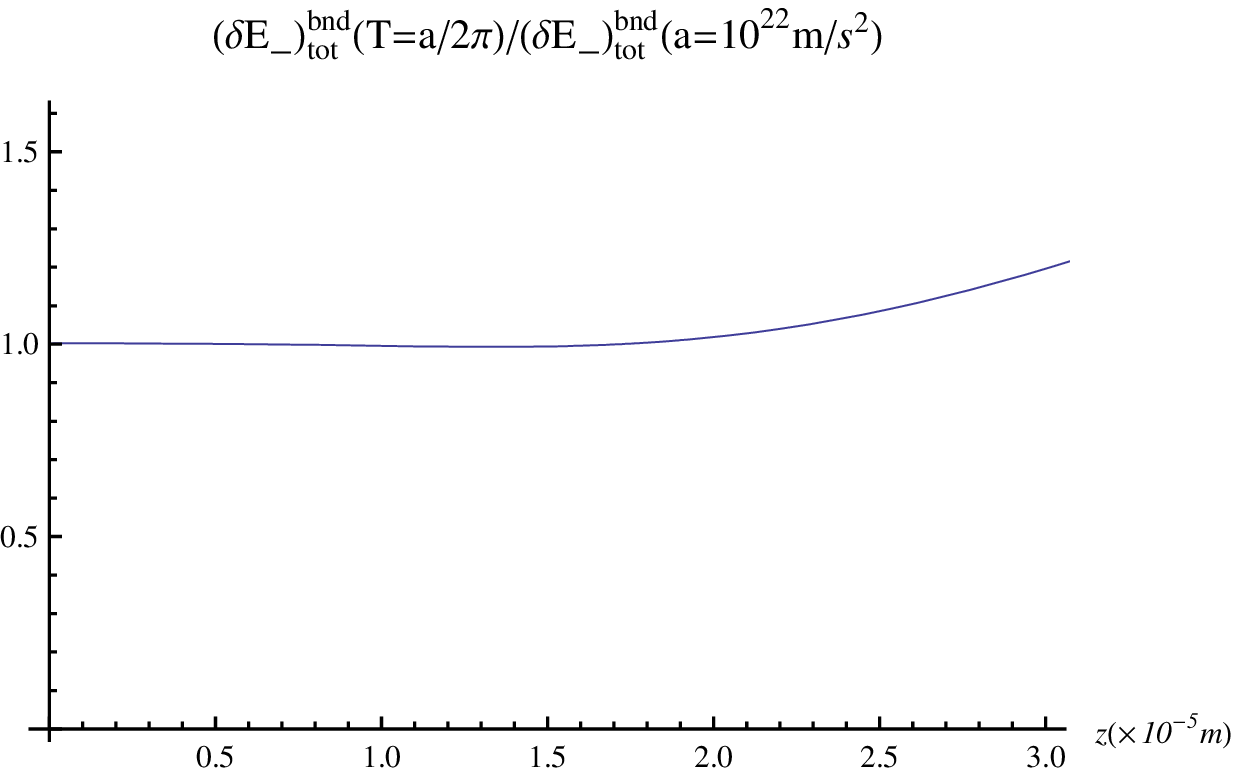}}
\caption{The ratio between the energy shift of a static atom
immersed in a thermal bath at the Unruh temperature $T=a/(2\pi)$ and
that of an accelerated one. Here two different values of
acceleration, $a=10^{23}$m/s$^2$ and $a=10^{22}$m/s$^2$, have been
considered.}\label{fig5}
\end{figure}

Finally, we  compare our results with those of a static atom
immersed in a thermal bath in the vicinity of an infinite conducting
plane \cite{Zhu-Yu09}. Take the acceleration to be
$a\sim10^{23}$\,m/s$^2$, and the corresponding Unruh temperature is
$T\sim405$\,K. We plot the energy shift of both an accelerated atom
and a static one immersed in a thermal bath at the Unruh temperature
in Fig.~\ref{fig4}, which reveals clearly that  the acceleration
effect is smaller than the thermal effect. So, the accelerated atom
does not behave as if immersed in a thermal bath at the Unruh
temperature in terms of the energy level shifts.
Like Ref.~\cite{Rizzuto09}, here we also consider the ratio between
the energy shift of an accelerated atom in the presence of an
infinite conducting plane and that of an atom at rest immersed in a
thermal bath at the corresponding Unruh temperature. In
Fig.~\ref{fig5}, we have plotted the quantity $(\delta
E_-)^{bnd}_{tot}(T={a\/2\pi})/(\delta E_-)^{bnd}_{tot}(a)$ as a
function of the distance $z$ for an isotropically polarized atom for
two different values of acceleration, $a\sim 10^{23}$m/s$^2$ which
satisfies $a\sim \omega_0$, and, $a\sim 10^{22}$m/s$^2$, which obeys
$a\ll \omega_0$.
These plots indicate that when the atom is very close to the
boundary, the ratio $(\delta E_-)^{bnd}_{tot}(T={a\/2\pi})/(\delta
E_-)^{bnd}_{tot}(a)$ approaches $1$ and is independent of the
acceleration of the atom. This is expected from our analytical
analysis, since in the short distance regime, $(\delta
E_-)^{bnd}_{tot}(T)$ agrees, in the leading order, with $(\delta
E_-)^{bnd}_{tot}(a)$  no matter  $a\sim \omega_0$, or  $a\ll
\omega_0$ (refer to Eq.~(\ref{z0}) and Eq.~(\ref{lowasmallz2})).

However, essentially, the effect of acceleration on the energy shift
 differs from that of the thermal one. As the distance increases,
this ratio grows.  The larger the acceleration, the more quickly the
ratio grows with the distance. This is also in contrast to the
conclusion drawn from Fig.~$4$ in Ref.~\cite{Rizzuto09}, where it is
found that the ratio may decrease and may become smaller than $1$
with the increase of the distance $z$.

\section{conclusions}
In conclusion, we have calculated separately the contributions of
vacuum fluctuations and radiation reaction to the position-dependent
energy shift of a uniformly accelerated atom interacting with
fluctuating vacuum electromagnetic fields in the vicinity of a plane
boundary, which gives rises to the Casimir-Polder force on the
accelerated atom. We have analyzed the behaviors of the energy level
shifts of the atom in different circumstances.

In the low-acceleration limit where the acceleration is much smaller
than the transition frequency of the atom, we found that, in the
short and intermediate distance regimes, the energy shift of the
accelerated atom is equal to that of a static one immersed in a
thermal bath at the Unruh temperature in the leading term. But in
the subleading term the acceleration corrections differ from the
thermal corrections at the Unruh temperature. In the long-distance
regime, even in the leading term, the behavior of an accelerated
atom   differs completely from that of the static one immersed in a
thermal bath at the Unruh temperature.

In the high-acceleration limit,  the acceleration correction is
equal, in the leading order,  to a thermal correction in the short
distance regime. However, the off-diagonal $xz$ component which is
absent in the thermal case makes the behavior of an accelerated atom
differ from that of a static atom immersed in a thermal bath at the
Unruh temperature. In the intermediate and long distance regimes,
the acceleration corrections completely differ from the thermal
corrections at the Unruh temperature.

For an acceleration of the order of the transition frequency of the
atom, we find, taking the frequency to be that of a hydrogen atom,
i.e., $\omega_0\sim10^{15}$\,s$^{-1}$,  that the effect of
acceleration  makes the energy shift smaller than that of an atom at
rest when the distance $z\lesssim10^{-3}$m, whereas when
$z\gtrsim10^{-3}$m, the energy shift oscillates significantly as the
distance increases. Therefore, there are some  distances  where the
effect of the acceleration on the energy shift is appreciable and
some other distances $z$ where the correction of acceleration
vanishes. It should be noted that although the effect of the
acceleration on the energy shift may in principle become
appreciable, it does not, however,  seem to be realistic in actual
measurements. Finally, compared with a static atom in a thermal
bath, we find that the energy shift of the accelerated atom close to
the boundary is smaller than that of the static one at the
corresponding Unruh temperature. It is worth pointing out that all
these features differ from what is found in Ref.~\cite{Rizzuto09}.

\begin{acknowledgments}

This work was supported in part by the National Natural Science
Foundation of China under Grants No. 10775050, No. 11075083,  No.
11005013 and No. 10935013; the Zhejiang Provincial Natural Science
Foundation of China under Grant No. Z6100077; the SRFDP under Grant
No. 20070542002; the National Basic Research Program of China under
Grant No. 2010CB832803; and the PCSIRT under Grant No. IRT0964.

\end{acknowledgments}

\end{document}